\documentclass[prl,aps,floatfix,superscriptaddress,footinbib,twocolumn,10pt,English]{revtex4}

\usepackage{amssymb,amsmath}
\usepackage{graphicx}
\usepackage[percent]{overpic}
\usepackage{overpic}
\usepackage{color}
\usepackage[colorlinks=true]{hyperref}
\usepackage{enumitem}
\usepackage{bm}

\begin{document}  

\title{Testing gravitational memory generation with compact binary mergers}

\author{ Huan Yang}
\affiliation{Perimeter Institute for Theoretical Physics, Waterloo, ON N2L2Y5, Canada}
\affiliation{University of Guelph, Guelph, ON N2L3G1, Canada}
\author{Denis Martynov}
\affiliation{LIGO, Massachusetts Institute of Technology, Cambridge, MA 02139, USA}
\affiliation{School of Physics and Astronomy and Institute of Gravitational Wave Astronomy, University of Birmingham, Edgbaston, Birmingham B15 2TT, United Kingdom}

\begin{abstract} 
Gravitational memory is an important prediction of classical General Relativity, which is intimately related to Bondi-Mezner-Sachs symmetries at null infinity
and the so-called soft graviton theorem first shown by Weinberg. For a given transient astronomical event, the angular distributions of energy and angular momentum flux
uniquely determine the displacement and spin memory effect in the sky. We investigate the possibility of using the binary black hole merger events detected by Advanced LIGO/Virgo to test the relation between  source energy emissions and  gravitational memory measured on earth, as predicted by General Relativity. We find that
while it is difficult for Advanced LIGO/Virgo, one-year detection of a third-generation detector network will easily rule out the hypothesis assuming isotropic memory distribution.
In addition, we have constructed a phenomenological model for memory waveforms of binary neutron star mergers, and use it to address the detectability of  memory from these events in the third-generation detector era. We find that measuring gravitational memory from neutron star mergers is a possible way to distinguish between different neutron star equations of state.
\end{abstract}

\maketitle 

{\noindent}{\bf Introduction}.~ With recent detection of binary neutron star (BNS) mergers  using both gravitational wave (GW) and electromagnetic telescopes \cite{PhysRevLett.119.161101,2041-8205-848-2-L12,2041-8205-848-2-L13},  
we are quickly entering the era of multi-messenger astronomy with GWs. Future GW observations will be able to provide unprecedented means
to uncover physical information of those most compact, exotic objects (such as black holes and neutron stars) in our universe. Moreover, future detections will open an independent window to study cosmology \cite{schutz1986determining,ligo2017gravitational},
 and will be used to test various predictions of General
Relativity \cite{yunes2016theoretical,berti2018extreme,berti2018extreme2}, such as the gravitational memory effect \cite{zel1974radiation,smarr1977gravitational,bontz1979spectrum,christodoulou1991nonlinear}. Gravitational memory itself is an observable  phenomenon of the spacetime, and conceptually it can be classified into ordinary memory originating from matter motions and GW memory  \footnote{Sometimes it is also called Christodoulou memory.} that arises from nonlinearities in the Einstein equation. The GW memory has a very intimate relation to soft-graviton
charges at null infinity \cite{he2015bms}, which may lead to quantum gravity partners  responsible for solving the Black Hole Information Paradox \cite{hawking2016soft}.
The latter possibility still contains significant uncertainty that requires further theoretical development \cite{mirbabayi2016dressed},
and it is unclear whether the memory effect is one of the few macroscopic, astrophysical observables that could be traced back to a quantum gravity origin (another example is ``echoes from black hole horizon" \cite{cardoso2016gravitational}).
Studying  such classical observables is interesting because  observation signatures of quantum gravity are normally expected  at  Planck scale.  
  
The detectability of the displacement memory  effect using ground, spaced-based detectors and pulsar-timing arrays has been discussed extensively in the literature \cite{thorne1992gravitational,lasky2016detecting,mcneill2017gravitational,favata2009nonlinear,favata2010gravitational,van2010gravitational,pollney2010gravitational}. 
 In addition, understanding and verifying the relation
between memory effect and associated energy/angular momentum emissions from the source is equally important, which displays striking similarities to Weinberg's soft-graviton theorem \cite{strominger2017lectures}.
Such relation has been written in various forms in different context.  In this work we adopt the form suitable to describe the nonlinear memory generated by GW energy flux \cite{thorne1992gravitational}:
\begin{align}\label{eqmem}
{h_{jk}^{\rm TT (mem)}}(T_d) =\frac{4}{d} \int^{T_d}_{-\infty} d t'\, \left [ \int \frac{d E^{\rm GW}}{dt' d \Omega'} \frac{n'_j n'_k}{1-{\bf n}' \cdot {\bf N}} d \Omega'\right ]^{\rm TT}\,,
\end{align}
 where $T_d$ is the time of detection, ${h_{jk}^{\rm TT (mem)}}$ is the memory part of the metric in transverse-traceless gauge, $\frac{d E^{\rm GW}}{dt' d \Omega'}$ is the GW energy flux, ${\bf n}'$ is its unit radial vector and
 ${\bf N}$ is the unit vector connecting the source and the observer (with distance $d$). 
   
 We propose to use binary black hole merger events  to test the validity of Eq.~\ref{eqmem}. 
 For any single event, a network of detectors is able to approximately determine its sky location and the intrinsic source parameters such as black hole masses, spins, and the orbital inclination,  
by applying parameter estimation algorithms. The displacement memory effect, being much weaker than the oscillatory part of the GW signal, can be also extracted using the matched-filter method.
 By computing GW energy with source parameters within the range determined by parameter estimation, we can obtain the value of the right-hand side of Eq.~\ref{eqmem} and compare it
 with measured displacement memory.
Multiple events are need to accumulate statistical significance for such a test \cite{yang2017black,yang2017gravitational}.

As an astrophysical application for gravitational memory, we also examine the memory generated by BNS mergers with a simple, semi-analytical memory waveform model. This memory waveform has a part that is sensitive to the star equation of state (EOS) and post-merger GW emissions. Therefore we are able to study the possibility of using memory detection to distinguish different NS EOS in the era of third-generation detectors.

\vspace{0.2cm}

{\noindent}{\bf Memory distribution}.~For binary black hole mergers at cosmological distances, the memory contribution can be well approximated by ($h^{\rm mem}_{\times}=0$ for circular orbit and standard choice of polarization basis) \cite{favata2009nonlinear,bieri2017gravitational} \footnote{This angular dependence assumes dominant (2,2) mode emission of GWs. For binary mergers with precessional spins, the effect from other mode emissions may also be included.}:
\begin{align}\label{eqa}
h^{\rm (mem)}_{+} =\frac{\eta M_z}{384 \pi d} \sin^2\iota (17+\cos^2\iota) h^{\rm mem}(T_d)\,,
\end{align}
where $M=m_1+m_2$ is the total mass of the binary, $z$ is the redshift, $M_z=M (1+z)$ is the redshifted total mass,  $\eta = m_1 m_2/M^2$ is the symmetric mass ratio, $\iota$ is the inclination angle of the orbit. The posterior distribution of these source parameters can be reconstructed by performing Markov-Chain Monte-Carlo parameter estimation procedure for each event.  $h^{\rm mem}$ can be well modelled by the {\rm minimal-waveform model} discussed in \cite{favata2009nonlinear}. The angular dependence shown in Eq.~\eqref{eqa} encodes critical information about memory generation described by Eq.~\eqref{eqmem}. It is maximized for edge-on binaries, which is different from the dominant oscillatory signals with $h_+ \propto (1+\cos^2\iota),\, h_\times \propto \cos \iota$ dependence. In this work, we test the consistency of Eq.~\eqref{eqa} with future GW detections as a way to test the memory generation formula Eq.~\eqref{eqmem}. In particular, we test the $\iota$-angle dependence \footnote{In principle we could also test the dependence of memory amplitude versus other source parameters, such the factor before $\sin^2\iota$ in Eq.~\eqref{eqa}. In the Bayesian model selection framework, such dependence can be compared to a null hypothesis, where the amplitude is zero, in which case it becomes a memory detection problem. We refer interested readers to \cite{lasky2016detecting} for related discussions.} and
  formulate this problem in a Bayesian model selection framework.
 
 \vspace{0.2cm}
 
 \begin{figure*}[tb]
\includegraphics[width=0.43\linewidth]{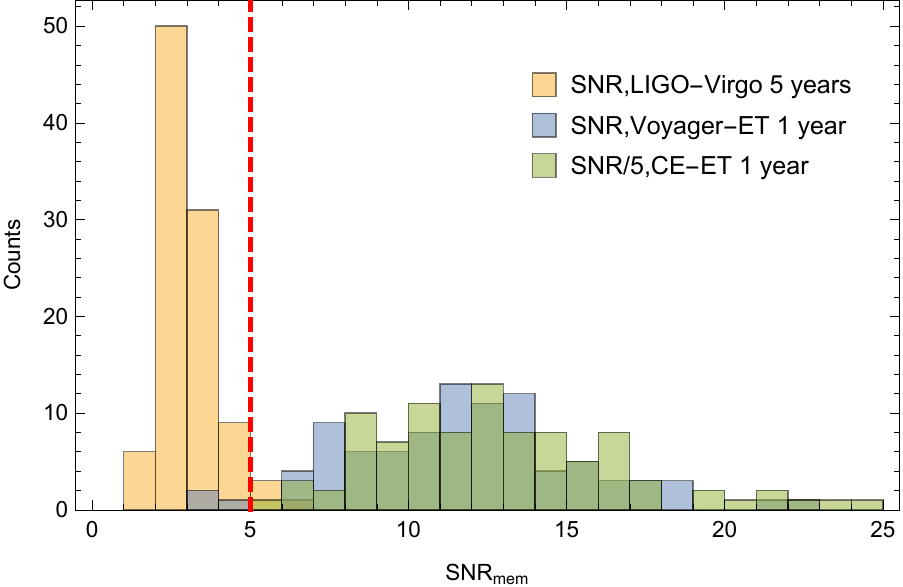}
\quad\quad
\includegraphics[width=0.43\linewidth]{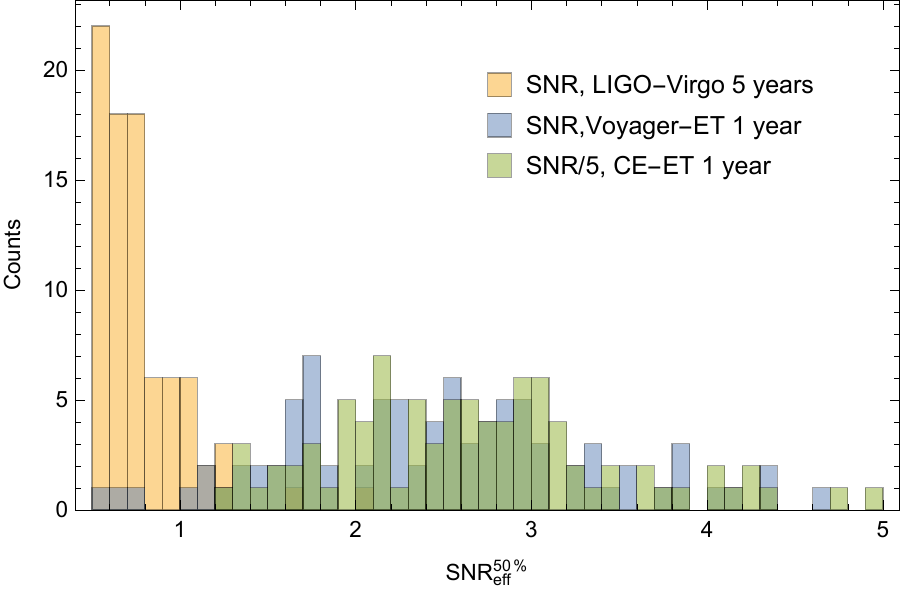}
\caption{Left panel:  The distribution of the combined SNR of the memory term for all events with expected
memory SNR greater than $0.1$, following A five year-observation with a network of GW detectors containing Advanced LIGO (Livingston and Hanford) and Advanced VIRGO
at design sensitivities. As a comparison, we also plot the combined SNRs for the same set of events assuming third-generation detectors. Right panel: The inferred ${\rm SNR}^{50\%}_{\rm eff}$ for distinguishing the two hypotheses in Eq.~\eqref{eqh1h2} for the
same set of detectors and with the same period of observation.}
\label{fig:config}
\end{figure*}

{\noindent}{\bf Model test}.~ We consider two following hypothesis, with $\mathcal{H}_1$ resembling Eq.~\eqref{eqa} and $\mathcal{H}_2$
describing an isotropic memory distribution in the source frame:
 \begin{align}\label{eqh1h2}
 \mathcal{H}_1: h^{\rm (mem)}_{+} & =\frac{\eta M_z}{384 \pi d} \sin^2\iota (17+\cos^2\iota) h^{\rm mem}(T_d)\,
 \equiv h_{m1}\,, \nonumber \\
\mathcal{H}_2: h^{\rm (mem)}_{+} & =\frac{\eta M_z}{96 \pi d} \sqrt{\frac{3086}{315}}h^{\rm mem}(T_d)\,
\equiv h_{m2}\,,
 \end{align}
 where the numerical coefficient of $h^{\rm (mem)}_{+}$ in $\mathcal{H}_2$ is chosen such that the (source) sky-averaged ${\rm SNR}^2$ (signal-to-noise ratio) is the
 same for these two hypothesis. For each detected  binary black hole merger event, the source parameters are described by
 
\begin{align}
\label{eq:par-ground}
\theta^a = (\ln \mathcal{M}_z, \ln  \eta, \chi, t_c, \phi_c, \ln d, \alpha,\delta,\psi, \iota)\,,
\end{align}
where $\mathcal{M}_z \equiv  M_z \eta^{3/5}$ is the redshifted chirp mass, $\chi \equiv (m_1 \chi_1+m_2 \chi_2)/M$ is the effective spin parameter~\cite{Ajith:2009bn} with $\chi_A$ representing the dimensionless spin of the $A$th body, $t_c$ and $\phi_c$ are the coalescence time and phase, $\alpha$, $\delta$ and $\psi$ are the right ascension, declination and polarization angle in the Earth fixed frame. Given a data stream $y$, to perform the hypothesis test, we evaluate the Bayes factor
\begin{align}
\mathcal{B}_{12} =\frac{P(\mathcal{H}_1 | y)}{P(\mathcal{H}_2 | y)}\,.
\end{align} 
In addition, the evidence $P(\mathcal{H}_i | y)$ is
\begin{align}\label{eqevidence}
P(\mathcal{H}_i | y) = \int d \theta^a P(\theta^a | \mathcal{H}_i) P(y | \theta^a \mathcal{H}_i)\,,
\end{align}
where the prior $P(\theta^a | \mathcal{H}_i) $ is  the prior distribution of $\theta^a$ which is set to be flat, and the likelihood function is given by
\begin{align}
 \log P(y | \theta^a \mathcal{H}_i) &\propto -2\int df \frac{| y-h_{\rm IMR}-  h_{mi} |^2}{S_n(f)} \,\nonumber \\
 & \equiv -\frac{||y-h_{\rm IMR}-  h_{mi}||^2}{2}\,,
 \end{align}
 with the inspiral-merger-ringdown waveform being $h_{\rm IMR}$ and the single-side detector noise spectrum $S_n$. Both $h_{\rm IMR}$ and $h_{mi}$ (cf. Eq.~\ref{eqh1h2}) are functions of $\{\theta^a\}$. According to the derivation in the Supplementary Material, after performing the integration in Eq.~\eqref{eqevidence}, the log of this Bayes factor can be approximated by
 \begin{align}\label{eqlogb12exp}
  \log  \mathcal{B}_{12} = & -\frac{1}{2} || y-h_{\rm IMR}(\hat{\theta})-  \epsilon h_{m1}(\hat{\theta}) ||^2 \nonumber \\
  &+\frac{1}{2} || y-h_{\rm IMR}(\hat{\theta})-\epsilon h_{m2}(\hat{\theta}) ||^2\,.
  \end{align}  
Here $\{ \hat{\theta}^a\}$ are the Maximum Likelihood Estimator for $\{ \theta^a\}$ using the IMR waveform template (PhenomB ~\cite{Ajith:2009bn} is adopted in this work). Similar to the discussion in \cite{Meidam:2014jpa,yang2017black,yang2017gravitational}, we denote the distribution of $  \log  \mathcal{B}_{12}$ in Eq.~\eqref{eqlogb12exp}  as {\it foreground} or {\it background} distributions, assuming hypothesis 1 or 2 is true respectively. Given a detected event, these  {\it foreground} and {\it background} distributions can be used to obtain the detection efficiency $P_{\rm d}$ and the false alarm rate $P_{\rm f}$ \cite{Meidam:2014jpa,yang2017black,yang2017gravitational}. Given an underlying set of source parameters $\theta_0=\{ \theta^a_0\}$, the false alarm rate can be obtained if the detection efficiency is known. In this work we follow the convention in \cite{abbott2017search} and choose $P_{\rm d} =50\%$.

For multiple events with data stream $\{ y^{(i)}\}$, the combined Bayes factor is 
\begin{align}
\mathcal{B}_{12} =\prod_i \frac{P(\mathcal{H}_1 | y^{(i)})}{P(\mathcal{H}_2 | y^{(i)})}\,,
\end{align}
and the above discussion generalizes trivially because these events are independent.
It turns out that, if we define ${\rm SNR}^{50\%}_{\rm eff}$ such that
\begin{align}
P^{50\%}_{\rm f} =\frac{1}{\sqrt{2\pi}} \int^\infty_{\rm SNR^{50\%}_{eff}} e^{-x^2/2} \,dx\,,
\end{align}
this effective SNR is  given by
\begin{align}
{\rm SNR^{50\%}_{eff}} = \frac{\sum_i || h^{(i)}_{m2}(\theta_0)-  h^{(i)}_{m1}(\theta_0) ||_i^2}{\sigma}\,,
\end{align}
with
\begin{align}\label{eqexp}
\sigma^2 & = \sum_i \left \{|| h^{(i)}_{m2}(\theta_0)-  h^{(i)}_{m1}(\theta_0) ||_i^2+A^{(i)}_a ({\Gamma^{(i)}_{ab}}^{-1}) A^{(i)}_b  \right \} \,,\nonumber \\
 \Gamma^{(i)}_{ab} & = \langle \partial_{\theta^a} h^{(i)}_{\rm IMR} | \partial_{\theta^b} h^{(i)}_{\rm IMR} \rangle_i \,,\nonumber \\
A^{(i)}_a & = \langle \partial_{\theta^a} h^{(i)}_{m1} | h^{(i)}_{m1}(\theta_0)-h^{(i)}_{m2}(\theta_0) \rangle_i \,,
\end{align}
and the inner product is defined as
  \begin{align}
 \langle \psi | \chi \rangle_i \equiv 2\int df \frac{\psi(f) \chi^*(f)+h.c.}{S_{n_i}(f)}\,.
 \end{align} 
The source parameter uncertainties enter into this hypothesis test result through the $A \Gamma^{-1} A$-type terms in Eq.~\ref{eqexp}. 
Because of the simplified treatment adopted in this analysis to save computational cost for simulated data, they are obtained essentially by the Fisher-Information method ($\Gamma$ is 
the Fisher-Information matrix). In principle, the whole procedure can also be performed using Markov-Chain Monte-Carlo method, where the posterior probability distribution of each parameter
can be more accurately computed.

\vspace{0.2cm}

{\noindent}{\bf Monte-Carlo source sampling}.~In order to investigate the distinguishability between different hypotheses over a given observation period, we randomly
sample merging binary black holes (BBHs) using a uniform rate in comoving volume $ 55{\rm Gpc}^{-3} {\rm yr}^{-1}$ consistent with \cite{abbott2016binary}.
The primary mass $m_1$ of the binary is sampled assuming a probability distribution $p(m_1) \propto m_1^{-2.35}$, where the secondary mass is uniformly sampled between $5 M_\odot$ and
$m_1$. We also require that the upper mass cut-off to be $M<80 M_\odot$ \cite{woosley2015deaths}.
The effective spin $\chi_i$ is sampled evenly within $|\chi_i| <1$.
The right ascension, declination, and inclination angles are randomly sampled assuming uniform distribution on the Earth's and source's sky.
We perform 100 Monte-Carlo realizations, each of which contains all BNS mergers within $z<0.5$ range (further binary merger events are too faint for memory detections) for a given observation period.

The results of the Monte-Carlo (MC) simulation are shown in Fig.~\ref{fig:config}. We assume a detector network with Advanced LIGO (both Livingston and Hanford sites) and Advanced Virgo, with all detectors reaching design sensitivity. After five-year observation time, we collect  all events with expected memory SNR above $0.1$ for each 
MC realization, and compute the corresponding ${\rm SNR}^{50\%}_{\rm eff}$ as defined in Eq.~\eqref{eqh1h2}. With a five-year observation, the median of this astrophysical distribution locates at $\sim 0.65 \sigma$ level, which is insufficient to claim a detection. 
Therefore under the current best estimate of merger rate and with the assumed binary BH mass distributions, during the operation period of Advanced LIGO-Virgo, it is unlikely to distinguish the (source) sky distribution of the memory term as depicted by Eq.~\eqref{eqmem}, \eqref{eqa} and an isotropic memory distribution.
In comparison, we apply the Voyager (or Cosmic Explorer, CE) sensitivity to both LIGO detectors, and the Einstein Telescope (ET) sensitivity to the Virgo detector,  and plot the corresponding SNR in 
Fig.~\ref{fig:config}. These 3rd-generation detector networks are fully capable of distinguishing the hypotheses.
Such a hypothesis test framework can also be applied to test against other memory distribution as well - one  needs to replace the second line of Eq.~\eqref{eqh1h2} by the target hypothesis. 

As an illustration, we also include the distribution of combined SNR: ${\rm SNR}_{\rm mem} = \sqrt{\sum_i ({\rm SNR}_{\rm mem}^{(i)})^2}$ \footnote{In order to coherently stack different data sets to boost the SNR of the stacked memory term, one needs to measure the high-order modes of the inspiral waveform to determine the signs of the memory terms in advance \cite{lasky2016detecting}. For hypothesis tests discussed in this work, such measurement is not required.}. This can be
achieved by adding the memory terms from different events coherently, as explained in \cite{lasky2016detecting}. 
 Its magnitude
roughly reflects the strength of combined memory signal over noise and the fact that its detection
is likely after five years' observation, which agrees with  \cite{lasky2016detecting}.

\vspace{0.2cm}

{\noindent}{\bf Recovering the angular dependence}. With a set of detections, it is also instructive to reconstruct the posterior angular dependence of memory, which can be compared with its theoretical prediction. Without loss of generality, we parametrize the memory waveform as
 \begin{align}\label{eqp}
h^{\rm (mem)}_{+} & =\frac{17 \eta M_z}{384 \pi d}  h^{\rm mem}(T_d) f(\{a_n, b_n\},\iota)\,, \nonumber \\
f(\{a_n, b_n\},\iota)& =\sum^N_{n=0} (a_n \sin n \iota + b_n \cos n \iota)\,,
 \end{align}
 where $N$ is the truncation wave number and $ h^{\rm mem}(T_d)$ is normalized to give the same Post-Newtonian waveform in the early inspiral stage. Given a set of observed events $y_j$, one can obtain the posterior distribution of  $a_i, b_i$ using  Bayes Theorem ($a_0 =0$):
 \begin{align}
 P(\{ a_i, b_i\} | \{ y_j\}) = \frac{P ( \{ y_j\} |\{ a_i, b_i\} ) P(\{ a_i, b_i\})}{P(\{ y_j\})}\,,
 \end{align}
 where the detailed expression for the likelihood function  $P ( \{ y_j\} |\{ a_i, b_i\} )$ is  explained in the Supplementary Material. In Fig.~\ref{fig:ang}, we simulate observed events (with ${\rm SNR}_{\rm m} \ge1$) in one year assuming CE-ET sensitivity. For simplicity, we assume that the memory distribution respects parity symmetry, such that all the $a_i$'s are zero. The cutoff $N$ is set to be $4$. Based on the posterior distribution of the angular distribution parameter $ b_i$, we compute the reconstructed uncertainty of $f_{\iota}$ at $1\sigma$ level, as depicted by the shaded area in Fig.~\ref{fig:ang}.
 
  \begin{figure}
\includegraphics[width=0.93\linewidth]{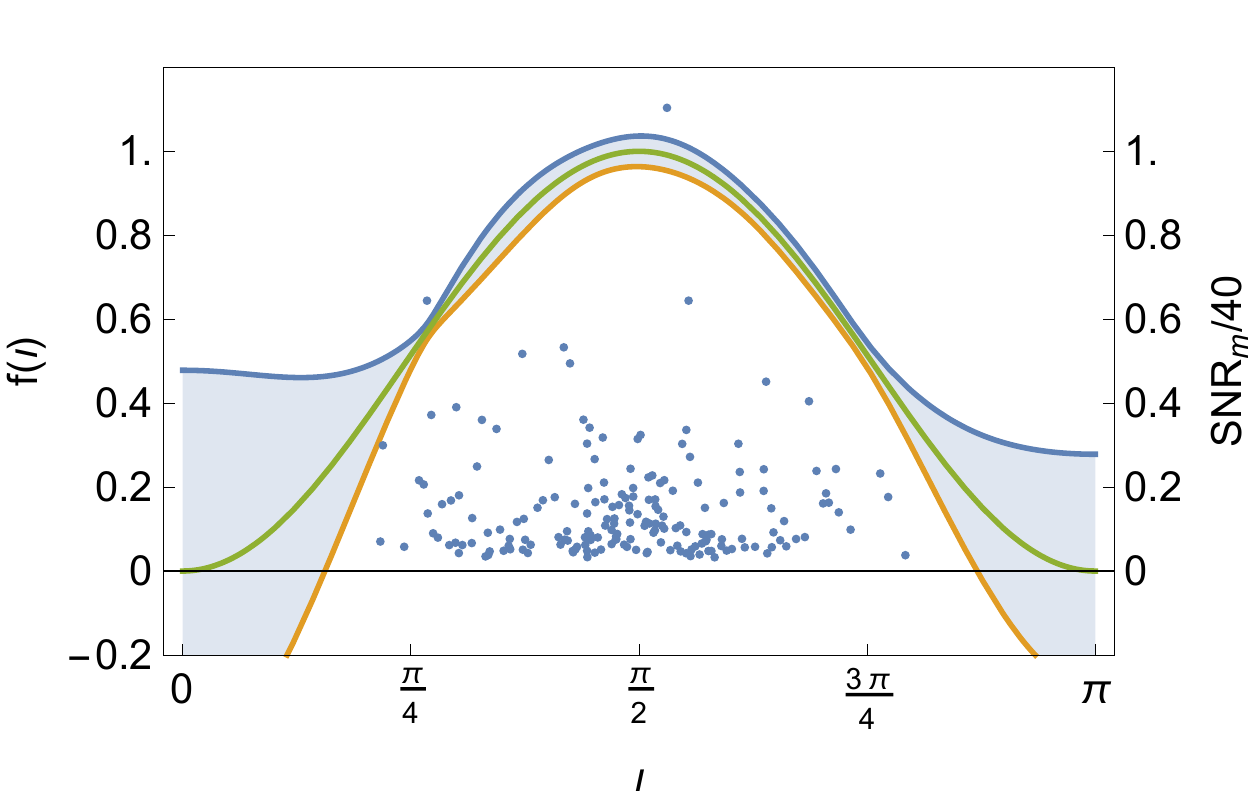}
\caption{The $1\sigma$ uncertainty of angular dependence $f(\iota)$ reconstructed from a set of simulated events, as indicated by the shaded region. The SNR and $\iota$ of simulated events are presented by the dots in the plot.}
\label{fig:ang}
\end{figure} 
 
\vspace{0.2cm}

 {\noindent}{\bf Binary neutron stars}.~ In addition to binary black holes, merging BNSs also generate a gravitational memory. However, as neutron star masses are smaller
 than the typical BH mass in binaries, and that the merger frequency is outside of the most sensitive band of current detectors, directly detecting gravitational memory from BNS mergers is difficult for second-generation detectors.  

Since the BNS waveform (especially the post-merger part) depends sensitively on the EOS, it is natural to expect that the detection of memory can be used to distinguish between various EOS. To achieve this goal, we have formulated a {\it minimal-waveform} model for BNS mergers similar to the construction for BBHs (see Supplementary Material). Such a model employs
the fitting formula for  post-merger waveforms developed in \cite{bose2017neutron} to compute $d E^{\rm GW}/dt$ (c.f. Eq.~\ref{eqmem}) in the post-merger stage, and a leading-PN description for the energy flux in the inspiral stage.
For illustration purpose, we also consider four sample EOS studied in  \cite{bose2017neutron}: GNH3, H4, ALF2, Sly. Assuming a $1.325M_\odot+1.325 M_\odot$ BNS system at distance 
$50{\rm Mpc}$ away from earth and following the maximally emitting direction, the SNRs for detecting these memory waveforms with Advanced LIGO are all around $0.1$, which are insufficient to study the EOS of neutron stars. On the other hand, if we assume Cosmic Explorer (CE) sensitivity, the corresponding SNRs will be 10.1, 9.6, 8.9, and 10.4 respectively.

For third-generation GW detectors such as CE, the inspiral waveform of BNS can be used to determine source parameters (such as $\iota$) to very high accuracies. For a $1.325M_\odot+1.325 M_\odot$ BNS system at distance $50{\rm Mpc}$ \footnote{Here we assume CE sensitivity for Handford, Livingston and Virgo detectors.}, Fisher analysis suggests that the measurement uncertainty of $\iota$ is of order $10^{-2}$. 
An accurate determination of source parameters breaks the degeneracy of amplitude between different BNS memory waveforms. We shall  compute 
\begin{align}
{\rm SNR}_{\Delta ab} = \sqrt{4 \int^\infty_0 df\,\frac{|\tilde{h}^{\rm mem}_{\rm MWM,a}-\tilde{h}^{\rm mem}_{\rm MWM,b}|^2}{S_{\rm n,CE}}}\,,
\end{align}
as a measure for distinguishability between  arbitrary EOS a and b. 

\begin{table}[b]
\caption{${\rm SNR}_\Delta$ for various EOS.}
\centering
\begin{tabular}{c c c c c}
\hline\hline
EOS & GNH3 & H4 & ALF2 & Sly  \\
\hline
GNH3 & 0 & 1.3 & 5.2 & 3.8  \\
H4 &  & 0 &  3.9 & 2.7\\
ALF2 &  &  & 0 & 2.3  \\
\hline
\end{tabular}
\label{table:bhnsdsnr}
\end{table}

According to the discussion in  \cite{lindblom2008model}, if  ${\rm SNR}_{\rm \Delta} \le 1$, we shall say that the two waveforms are indistinguishable.
The values listed in Table \ref{table:bhnsdsnr} indicate that measuring gravitational memory  is a possible way to extract information about
neutron star EOS. One unique advantage of this approach is that it is insensitive to phase difference between post-merger modes, as the beating term between
modes generally contribute $k$Hz modulation of $d E^{\rm GW}/dt$ or $h^{\rm mem}$, which is outside the most sensitive band of third-generation detectors \footnote{Unless it is a high-frequency detector targeting $k$Hz band, such as the one discussed in \cite{miao2017towards}.}. Such mode phases still contain much more significant theoretical uncertainties than mode frequencies in current numerical simulations.

\vspace{0.2cm}

 {\noindent}{\bf Memory for ejecta}.~The electromagnetic observation of GW170817 provides strong evidence for multi-component ejecta \cite{hallinan2017radio,smartt2017kilonova}, which could originate from collisions of stars, wind from post-collapse disk \cite{siegel2017three}, etc. Because of the transient nature, the GWs generated by ejecta(s) are likely non-oscillatory, which are mainly composed by ordinary gravitational memory \cite{braginsky1987gravitational}:
 \begin{align}\label{eqlmem}
h_{jk}^{\rm TT (mem)} = \Delta \sum^N_{A=1} \frac{4 M_A}{d \sqrt{1-v_A^2}} \left [ \frac{v^j_A v^k_A}{1-{\bf v_A} \cdot {\bf N}}\right ]^{\rm TT}\,.
\end{align}
We shall phenomenologically write the ejecta waveform as $h_+ = h_0 (1+e^{-t/\tau})^{-1}$, with the frequency domain waveform being $i \pi \tau /\sinh(2\pi^2 f \tau)$. Here $\tau$ characterizes the duration of the ejection process, and $h_0$ is the asymptotic magnitude of the linear memory. Depending on the angular distribution of the ejecta material, $h_0$ along the maximally emitting direction can be estimated as $h_0 \sim \Delta M v^2/d$, where $\Delta M$ is the ejecta mass and $v$ is the characteristic speed.
Assuming CE sensitivity, the SNR of such ejecta waveforms is a plateau for $\tau \le 1$ms, and drops quickly for larger $\tau$. The plateau value roughly scales as \footnote{We assume that the lower cut-off frequency for computing SNR is $5$Hz.}
\begin{align}
{\rm SNR}_{\rm ej} \sim 1.2 \left ( \frac{\Delta M}{0.03 M_\odot} \right ) \left ( \frac{v}{0.3 c} \right )^2 \left ( \frac{d}{50 {\rm Mpc}} \right )^{-1}\,.
\end{align}  

 In this case, a detection of ejecta waveforms is only plausible  
with information stacked from multiple events, and/or using detectors that
achieve better low frequency sensitivity \cite{yu2017prospects}.
Out of curiosity, one can apply a similar analysis to the jet of a short gamma-ray burst.
The SNR roughly scales as $\sim 0.25 (\Delta E_{\rm jet}/10^{51} {\rm erg}) (50 {\rm Mpc}/d)$, which is even smaller.

\vspace{0.2cm}

{\noindent}{\bf Conclusion}.~ We have discussed two aspects of measuring gravitational memory in merging compact binary systems. For BBHs, it is ideal to test the memory-generation mechanism, as a way to connect  soft-graviton theorem and symmetry charges of the spacetime to astrophysical observables. For BNSs, it can be used to distinguish between different NS EOS, as a complementary way to tidal love number measurements in the inspiral waveform and (possibly) spectroscopy measurements of the post-merger signal. We have shown that both tasks may be achieved with the third-generation detectors.  
 
 Because of the $1/f$-type scaling of memory waveforms, improving the low-frequency sensitivity of detectors is crucial for achieving better memory SNR. This will be particularly useful for gravitationally probing the ejecta(s)  produced in BNS mergers. Another interesting direction will be further exploring the detectability and application of memory in space-based missions, such as LISA or DECIGO.

 {\bf Acknowledgments.}
We would like to thank 
Haixing Miao and Lydia Brieri for 
fruitful discussions.
We thank Yuri Levin for reading over the manuscript and making many useful comments.
H.Y. is supported in
part by Perimeter Institute for Theoretical Physics. Research at
Perimeter Institute is supported by the Government of Canada through
Industry Canada and by the Province of Ontario through the Ministry
of Research and Innovation. 
D.M. acknowledge the support of the NSF and the Kavli Foundation.

 \appendix
  
 \section{Details of the hypothesis test} 
 
 We are testing two hypotheses:
 \begin{align}
 \mathcal{H}_1: y  & =\delta h_{\rm IMR}+n+ \epsilon \frac{\eta M_z}{384 \pi d} \sin^2\iota (17+\cos^2\iota) h^{\rm mem}(T_d)\,,\nonumber \\
 & = \delta h_{\rm IMR}+n + \epsilon h_{m1} \nonumber \\
\mathcal{H}_2: y & =\delta h_{\rm IMR}+n+\frac{\eta M_z}{96 \pi d} \sqrt{\frac{3086}{315}}h^{\rm mem}(T_d)\, \nonumber \\
 & = \delta h_{\rm IMR}+n + \epsilon h_{m2} \,,
 \end{align} 
 where $\epsilon$ is a book-keeping parameter to track the power of the memory terms (as they are generally smaller than the oscillatory part), $n$ is the detector noise, $\delta h_{\rm IMR}$ is the residual part due to imperfect subtraction of the oscillatory part of the inspiral-merger-ringdown (IMR) waveform. Notice that the overlap between the memory waveform and the IMR is very small. For example, for GW150914-like events, the overlap is
 \begin{align}
 \mathcal{F}(h^{\rm mem}, h_{\rm IMR}) =\frac{\langle h^{\rm mem} | h_{\rm IMR} \rangle}{\sqrt{\langle h^{\rm mem} | h^{\rm mem}\rangle} \sqrt{\langle h_{\rm IMR} | h_{\rm IMR}\rangle}} \approx 0.7\%\,,
 \end{align}
 where the inner product is
 \begin{align}
 \langle \psi | \chi \rangle \equiv 2\int df \frac{\psi(f) \chi^*(f)+h.c.}{S_n(f)}\,, 
 \end{align} 
  where $S_n(f)$ is the single-side detector spectrum. Similarly we can check $\mathcal{F}(h^{\rm mem}, \partial_{\theta^a} h_{\rm IMR})$ are of similar order. 
  As a result, we approximate the memory waveform to be orthogonal to the oscillatory part of the waveform.

  Consider 
  
 \begin{align}
P(\mathcal{H}_i | y) = \int d \theta^a P(\theta^a | \mathcal{H}_i) P(y | \theta^a \mathcal{H}_i)\,,
\end{align}
  where the prior $P(\theta^a | \mathcal{H}_i)$ is taken to be flat. On the other hand, the likelihood function $P(y | \theta^a \mathcal{H}_i)$ is  given by
  \begin{align}
  P(y | \theta^a \mathcal{H}_i) \propto e^{-1/2 \langle y-h_{\rm IMR}- \epsilon h_{mi} | y-h_{\rm IMR}- \epsilon h_{mi} \rangle}\,,
  \end{align}
  such that
  \begin{align}\label{eqlikeli}
P(\mathcal{H}_i | y) = \int d \theta^a e^{-1/2 || y-h_{\rm IMR}- \epsilon h_{mi} ||^2 }\,.
\end{align}
Here both $h_{\rm IMR}$ and $h_{mi}$ are functions of $\theta^a$. We further choose the $\hat{\theta}^a$ such that 
\begin{align}
\left . \langle y | \partial_{\theta^a} h_{\rm IMR} \rangle \right |_{\hat{\theta}^a} = \left . \langle h_{\rm IMR} |  \partial_{\theta^a} h_{\rm IMR} \rangle \right |_{\hat{\theta}^a}\,.
\end{align}
 In other words, $\hat{\theta}^a$ are the Maximum Likelihood Estimators of $\theta^a$ using the matched filter $h_{\rm IMR}$. We can further expand the exponent of 
  Eq.~\eqref{eqlikeli} to be
  \begin{align}
  &y-h_{\rm IMR}- \epsilon h_{mi} \nonumber \\
  &\approx y-h_{\rm IMR}(\hat{\theta}^a)- \epsilon h_{mi}(\hat{\theta}^a) -\partial_{\theta^a} h_{\rm IMR} \delta \theta^a-\epsilon \partial_{\theta^a} h_{mi} \delta \theta^a\,,
  \end{align}
  where $\delta \theta^a =\theta^a-\hat{\theta}^a\,$. By applying the orthogonality condition between the IMR waveform and memory waveform and removing terms at $\mathcal{O}(\epsilon^2)$
  order, after the Gaussian integration in Eq.~\eqref{eqlikeli} we find that 
   \begin{align}\label{eqlikeli2}
P(\mathcal{H}_i | y) \propto  e^{-1/2 || y-h_{\rm IMR}(\hat{\theta})- \epsilon h_{mi}(\hat{\theta}) ||^2 } \frac{1}{\sqrt {{\rm det}(\Gamma_{ab})}}\,.
\end{align}
  with
  \begin{align}
  \Gamma_{ab} = \langle \partial_{\theta^a} h_{\rm IMR} | \partial_{\theta^b} h_{\rm IMR} \rangle\,.
  \end{align}
  As a result, the log Bayes factor is given by
  \begin{align}
  \log  \mathcal{B}_{12} = & -\frac{1}{2} || y-h_{\rm IMR}(\hat{\theta})-  \epsilon h_{m1}(\hat{\theta}) ||^2 \nonumber \\
  &+\frac{1}{2} || y-h_{\rm IMR}(\hat{\theta})-\epsilon h_{m2}(\hat{\theta}) ||^2\,.
  \end{align}

With underlying source parameters $\theta^a_0$ and assuming hypothesis $2$ is true, we can evaluate the {\it background distribution} of the log Bayes factor \cite{Meidam:2014jpa,yang2017black,yang2017gravitational}. Let us denote 
\begin{align}
s & =y -h_{\rm IMR}(\hat{\theta})- \epsilon h_{m2}(\hat{\theta})=n +[h_{\rm IMR}(\theta_0)-h_{\rm IMR}(\hat{\theta})] \nonumber \\
&+\epsilon [h_{\rm m2}(\theta_0)-h_{\rm m2}(\hat{\theta})]
 = n + \delta h_{\rm IMR} +\epsilon \delta h_{m2}\,.
\end{align}
The log Bayes factor becomes 
  \begin{align}\label{eqlogb1}
  \log  \mathcal{B}_{12} = & -\frac{1}{2} \epsilon^2 || h_{m2}(\hat{\theta})-  h_{m1}(\hat{\theta}) ||^2 \nonumber \\
 &  + \epsilon \langle s | h_{m1}(\hat{\theta})-h_{m2}(\hat{\theta}) \rangle\, \nonumber \\
 & = -\frac{1}{2} \epsilon^2 || h_{m2}(\hat{\theta})-  h_{m1}(\hat{\theta}) ||^2 \nonumber \\
 &  + \epsilon \langle n+\epsilon \delta h_{m2} | h_{m1}(\hat{\theta})-h_{m2}(\hat{\theta}) \rangle\, \nonumber \\
 & = -\frac{1}{2} \epsilon^2 || h_{m2}(\theta_0)-  h_{m1}(\theta_0) ||^2 \nonumber \\
 &  + \epsilon \langle n+\epsilon \delta h_{m1} | h_{m1}(\theta_0)-h_{m2}(\theta_0) \rangle\,\nonumber \\
 & \approx -\frac{1}{2} \epsilon^2 || h_{m2}(\theta_0)-  h_{m1}(\theta_0) ||^2 \nonumber \\
 &  + \epsilon \langle n | h_{m1}(\theta_0)-h_{m2}(\theta_0) \rangle \nonumber \\
 &+\epsilon^2 \delta \theta^a_0 \langle \partial_{\theta^a} h_{m1} | h_{m1}(\theta_0)-h_{m2}(\theta_0) \rangle\,.
  \end{align} 
Notice that if we normalize the magnitude of $\langle n | h_{\rm IMR} \rangle/||h_{\rm IMR}||$  or  $\langle n | h_{mi} \rangle/||h_{mi}||$ as $\sim 1$ , we have $\delta \theta^a_0 = \theta^a_0 -\hat{\theta}^a \sim 1/{\rm SNR}_{\rm IMR}$ and $|| h_{mi} || \sim {\rm SNR}_{\rm mem}$.  That's why we have dropped terms like $\delta \theta^a_0 \delta \theta^b_0 \langle \partial_a h_{mi} | \partial_b h_{mi} \rangle \sim (\rm SNR_{mem}/SNR_{IMR})^2$. Let us denote the distribution of the last three lines of Eq.~\eqref{eqlogb1} as $P_1$, the false alarm
probability (rate) of a given detection is 
\begin{align}
P_{\rm f} = \int^\infty_{\log \mathcal{B}_{12}} P_1(X) dX \equiv R_1(\log \mathcal{B}_{12})\,.
\end{align}

On the other hand, assuming hypothesis $1$ is true, the log Bayes factor becomes
 \begin{align}\label{eqlogb2}
  \log  \mathcal{B}_{12} 
 & \approx \frac{1}{2} \epsilon^2 || h_{m2}(\theta_0^a)-  h_{m1}(\theta_0^a) ||^2 \nonumber \\
 &  + \epsilon \langle n | h_{m1}(\theta_0^a)-h_{m2}(\theta_0^a) \rangle \nonumber \\
 &+\epsilon^2\delta \theta^a_0 \langle \partial_{\theta^a} h_{m2} | h_{m1}(\theta_0^a)-h_{m2}(\theta_0^a) \rangle\,.
  \end{align}
 Let us denote the distribution of the last three lines of Eq.~\eqref{eqlogb2} as $P_2$, the detection efficiency (probability) is 
 \begin{align}
P_{\rm d} = \int^\infty_{\log \mathcal{B}_{12}} P_d(X) dX \equiv R_2(\log \mathcal{B}_{12})\,.
\end{align}

For a given detection efficiency (say $50\%$), we can obtain the false alarm probability $P^{50\%}_{\rm f}$ based on the underlying source parameter $\theta^a_0$. Such a false
alarm rate can be mapped to an effective SNR of a standard Gaussian distribution:
\begin{align}
P^{50\%}_{\rm f} =\frac{1}{\sqrt{2\pi}} \int^\infty_{\rm SNR^{50\%}_{eff}} e^{-x^2/2} \,dx\,.
\end{align}

According to the set up of this problem, one can show that $\rm SNR^{50\%}_{eff}$ is 
\begin{align}
{\rm SNR^{50\%}_{eff}} = \frac{|| h_{m2}(\theta_0)-  h_{m1}(\theta_0) ||^2}{\sigma}\,,
\end{align}
with
\begin{align}
\sigma^2 & = || h_{m2}(\theta_0)-  h_{m1}(\theta_0) ||^2+A_a (\Gamma_{ab}^{-1}) A_b \nonumber \\
A_a & = \langle \partial_{\theta^a} h_{m1} | h_{m1}(\theta_0)-h_{m2}(\theta_0) \rangle\,.
\end{align}
  
\section{Angular dependence recovery}
  
With a set of observations $y_i$,  we first cross product each data stream with the memory waveform $h^{\rm m}(\Theta,T_d) \equiv \frac{17 \eta M_z}{384 \pi d}  h^{\rm mem}(T_d)$, with $\Theta$ being a generalization of $\theta^a$ which includes individual spins. 
\begin{align}
s_i \equiv \langle y_i | h^{\rm m,i}(\Theta_i) \rangle f(\{a_n, b_n\},\iota) \rangle\,.
\end{align}
 The likelihood function $\mathcal{L}(s_i | \{ a_n, b_n\}, \Theta_i)$ is
 \begin{align}\label{eqlike}
& \mathcal{L}(s_i | \{ a_n, b_n \}, \Theta_i) \propto \nonumber \\
&{\rm exp} \left [ -\frac{ (s_i - ||h^{\rm m}||^2 f(\{a_n, b_n\},\iota) )^2}{2 ||h^{\rm m}||^2}\right ]\,.
 \end{align} 
 
For the events we are considering here, the SNR of the oscillatory part of the waveform is roughly $50-100$ times larger than the SNR of the memory waveform. Third-generation detectors are generally required for performing the angular dependence recovery of memory. As a result, $\Theta_i$ can be assumed to be accurately determined  (with posterior distribution $\pi$) from the oscillatory part of $y_i$, such that
\begin{align}
& \int d \Theta_i \mathcal{L}(s_i | \{ a_n, b_n \}, \Theta_i) \pi(\Theta_i) \nonumber \\
& \approx \mathcal{L}(s_i | \{ a_n, b_n \}, \hat{\Theta}_i)\,,
\end{align} 
where $\hat{\Theta}_i$ are the Maximum Likelihood Estimators for $\Theta_i$. According to  Bayes' Theorem, the posterior distribution of $\{ a,b\}$ is
\begin{align}
P(\{ a_n, b_n\} | \{ y_j\}) \propto \prod_i  \mathcal{L}(s_i | \{ a_n, b_n \}, \hat{\Theta}_i)\,.
\end{align}
 Based on the function form of Eq.~\ref{eqlike}, the distribution of $\{ a_n, b_n\}$ is still Gaussian, with variance matrix given by
 \begin{align}
 & V^{-1}_{a_n, b_l} = \sum_i || h^{\rm m,i}||^2 \cos n \iota_i \sin l \iota_i\,,\nonumber \\
  & V^{-1}_{a_n, a_l} = \sum_i || h^{\rm m,i}||^2 \cos n \iota_i \cos l \iota_i\,, \nonumber \\
   & V^{-1}_{b_n, b_l} = \sum_i || h^{\rm m,i}||^2 \sin n \iota_i \sin l \iota_i\,.
 \end{align}
  
 \section{Memory waveform for binary neutron star mergers} 
   
  \begin{widetext} 
   
\begin{table}[b]
\caption{Parameters for various EOS \cite{bose2017neutron}}
\centering
\begin{tabular}{c c c c c c c c c c}
\hline\hline
EOS & $f_1$ ($k$Hz) & $\tau_1$ (ms) & $f_2$ ($k$Hz) & $\tau_2$ (ms)  & $\gamma_2$ (${\rm Hz}^2$) & $\xi_2$ (${\rm Hz}^3$) & $\alpha$ & $r_m$ (km) & A (km)\\
\hline
GNH3 & 1.7 & 2 & 2.45 & 23.45 & 342 & 5e4 & 0.35 & 28.2 & 0.726 \\
H4 & 1.75 & 5 &  2.47 & 20.45 & -1077 &  4.5e3 & 0.3 & 27.5 & 0.692\\
ALF2 & 2.05 & 15 & 2.64 & 10.37 &  -863 & 2.5e4 & 0.5 & 26 & 0.519 \\
Sly & 2.3 & 1 & 3.22 & 13.59 & -617  & 5.5e4 & 0.5 & 24.7 & 0.554 \\
\hline
\end{tabular}
\label{table:bhnssnr}
\end{table}

\end{widetext}   
   
We shall construct an analytical memory waveform model for binary neutron star mergers similar to the approach adopted in \cite{favata2009nonlinear} for binary black holes.
Following the {\it minimal-waveform model}, the memory waveform can be computed using the radiative moment
\begin{align}
h^{\rm mem}(T_d) = \frac{1}{\eta M} \int^{T_d}_{-\infty} |I^{(3)}_{22}(t)|^2 \, dt\,.
\end{align}
We match the leading order inspiral moment to the moment of post-merger hypermassive neutron stars. The qth derivative of the inspiral moment is given by
\begin{align}\label{eqi22}
I^{\rm insp(q)}_{22} =2\sqrt{\frac{2 \pi}{5}} \eta M r^2 (-2 i \omega)^q e^{-2 i \phi}\,,
\end{align}
where $\phi$ is the 0PN orbital phase, $\omega=\dot{\phi}=(M/r^3)^{1/2}$, $r=r_m (1-T/\tau_{rr})^{1/4}$ is the orbital separation, $T=t-t_m$ is the time since the matching time $t_m$,  $\tau_{rr}=(5/256)(M/\eta)(r_m/M)^4$, and $r_m$ is the distance at the matching time.  On the other hand, because $h^{TT}_{ij} \sim \ddot{I}_{ij}/d$ and the post-merger waveform can be approximately parametrized as \cite{bose2017neutron}
\begin{align}
h_{\rm post}(t) \propto & \,\alpha 
e^{-t/\tau_1}[\sin 2\pi f_1 t +\sin 2 \pi (f_1-f_{1\epsilon}) t \nonumber \\
& +  \sin 2 \pi (f_1+f_{1\epsilon} )t ] \nonumber \\
&+e^{-t/\tau_2} \sin (2\pi f_2 t+2 \pi \gamma_2 t^2+2 \pi \xi_2 t^3+\pi \beta_2)\,,
\end{align}
with the waveform parameters given in Table.~\ref{table:bhnssnr} for various EOS considered here,
we find that
\begin{align}
I^{\rm post(2)}_{22} = & -i A \alpha e^{-T/\tau_1}[e^{2\pi i f_1 T} +e^{ 2 \pi i (f_1-f_{1\epsilon} ) T} \nonumber \\
&+ e^{ 2 \pi i (f_1+f_{1\epsilon} )T} ] \nonumber \\
&- i A e^{-T/\tau_2} e^{ 2\pi i f_2 T+2 \pi i \gamma_2 T^2+2 \pi i \xi_2 T^3+i \pi \beta_2}\,,
\end{align}
with A determined by fitting with the numerical post-merger waveform. In the timescale of interest ($\tau_1$ or $\tau_2$), we have $f_2, f_1 \gg \gamma_2 \tau \& \xi_2 \tau^2$.
Therefore we shall simplify $I^{\rm post(2)}_{22}$ to be
\begin{align}
I^{\rm post(2)}_{22} \approx & -i A \alpha e^{-T/\tau_1}[e^{2\pi i f_1 T} +e^{ 2 \pi i (f_1-f_{1\epsilon} )T} + e^{ 2 \pi i (f_1+f_{1\epsilon} )T} ] \nonumber \\
&- i A e^{-T/\tau_2} e^{ 2\pi i f_2 T+i \pi \beta_2}\,\nonumber \\
& =\sum^4_{i=1} A_i e^{i 2 \pi f_i t-t/\tau_i}
\end{align}

 \begin{figure}[tb]
\includegraphics[width=0.93\linewidth]{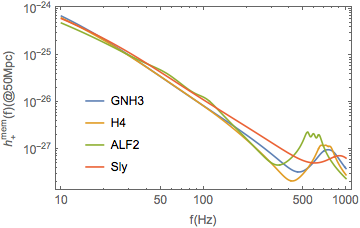}
\caption{Gravitational memory waveforms  for a $1.325 M_\odot +1.325 M_\odot$ binary neutron star system at $50$Mpc, assuming different EOS and along the maximally emitting direction.}
\label{fig:ej}
\end{figure} 

 \begin{figure}[tb]
\includegraphics[width=0.93\linewidth]{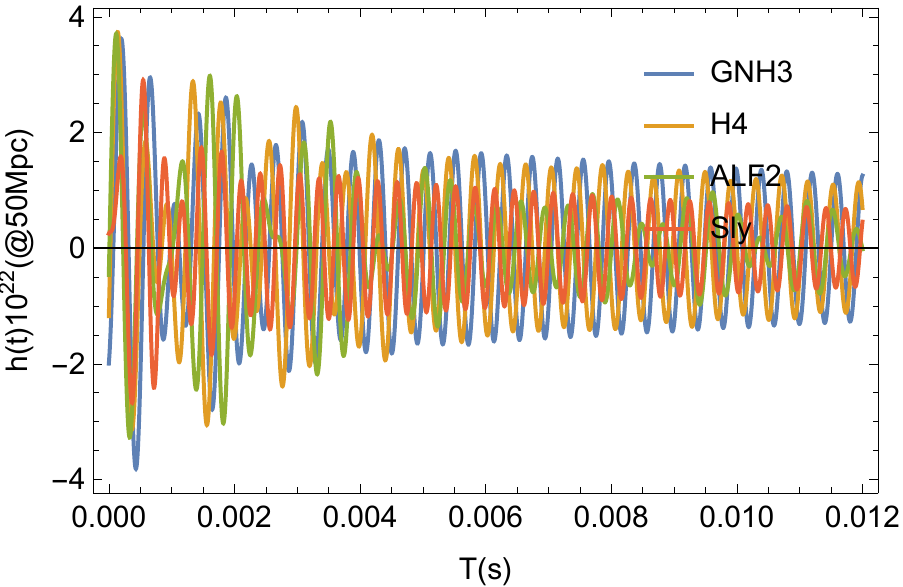}
\caption{Post-merger waveforms for $1.325 M_\odot+1.325 M_\odot$ binary neutron star system with four EOS considered in this work (GNH3, H4, ALF2, Sly). The distance is assumed to be $50$Mpc.}
\label{fig:wave}
\end{figure}

The 2nd derivative of inspiral and post-merger radiative moments are matched at $t_m$, which has the physical meaning of continuity of $h$. 
$r_m$ can be estimated by twice the radius of the stars. An alternative way to fix $r_m$ is to use the oscillation amplitude of $h_+$ right before merger \cite{favata2009gravitational}:
\begin{align}
h_+ -i h_\times \approx &\frac{1}{8 d}\sqrt{\frac{5}{2 \pi}} \left [ (1+\cos \iota)^2 e^{2 i \Phi} I^{(2)}_{22} \right . \nonumber \\
& \left .+(1-\cos\iota)^2 e^{-2 i\Phi} I^{(2)}_{2-2}\right ]\,,
\end{align}
with $\Phi$ being the direction of the observer in the source frame. Combining with Eq.~\eqref{eqi22}, we find that the amplitude of $h_+$ along the maximum emitting direction is
\begin{align}
h_{+m} = \frac{4 \eta M r^2 \omega^2}{d}\,,
\end{align}
which can be used to determine $r_m$.  Similarly, it is straightforward to obtain that
\begin{align}
A_{22,+m}(T=0) = \frac{1}{2 d} \sqrt{\frac{5}{2 \pi}} A\,,
\end{align}
where $A_{22}$ is the amplitude of $22$ mode. This can be used to determine $A$.

The memory waveform in the time domain is ($\sigma_i \equiv i 2 \pi f_i +\tau^{-1}_i$)
\begin{align}
h^{mem}_{MWM}(T_d) \approx & \frac{8 \pi M}{ r(T_d)} \Theta(-T_d) +\Theta(T_d) \left \{ \frac{8 \pi M}{ r_m} +\frac{1}{ \eta M} \right . \nonumber \\
& \left. \sum^4_{i,j=1} \frac{\sigma_i \sigma^*_j A_i A^*_j}{\sigma_i+\sigma_j} [1-e^{-(\sigma_i+\sigma^*_j)t}]\right \}\,,
\end{align}
where $\Theta(T_d)$ is the Heaviside function. The corresponding frequency domain waveform is
\begin{align}
\tilde{h}^{mem}_{MWM}(f) = & \frac{i}{2 \pi f} \left \{ \frac{8 \pi M}{r_m} [1-2 \pi i f \tau_{rr} U(1,7/4, 2\pi i f \tau_{rr})] \right . \nonumber \\
& \left. -\frac{1}{\eta M} \sum^4_{i,j=1} \frac{\sigma_i \sigma^*_j A_i A^*_j}{2 \pi i f -(\sigma_i +\sigma^*_j)} \right \}\,,
\end{align}
where U is Kummer's confluent hypergeometric function of the second kind. The high frequency poles above $1$kHz are unimportant for the analysis assuming ET or CE, because their low-frequency sensitivity is superior compared to their high frequency sensitivity.

\section{SNR of the ejecta waveform}

Following the discussion in the main text, we assume the memory waveform model to be
\begin{align}
\tilde{h}^{\rm mem}_{\rm MWM,m} = \frac{\Delta M v^2}{d} \frac{i \pi \tau }{\sinh(2\pi^2 f \tau)}\,.
\end{align}

We show the corresponding SNR as a function of $\tau$ in Fig.~\ref{fig:ej}. We find that for $\tau \le 1$ms, the SNR is relatively flat $\sim 1.2$. For larger $\tau$ values, the SNRs also decrease dramatically. 

 \begin{figure}[tb]
\includegraphics[width=0.93\linewidth]{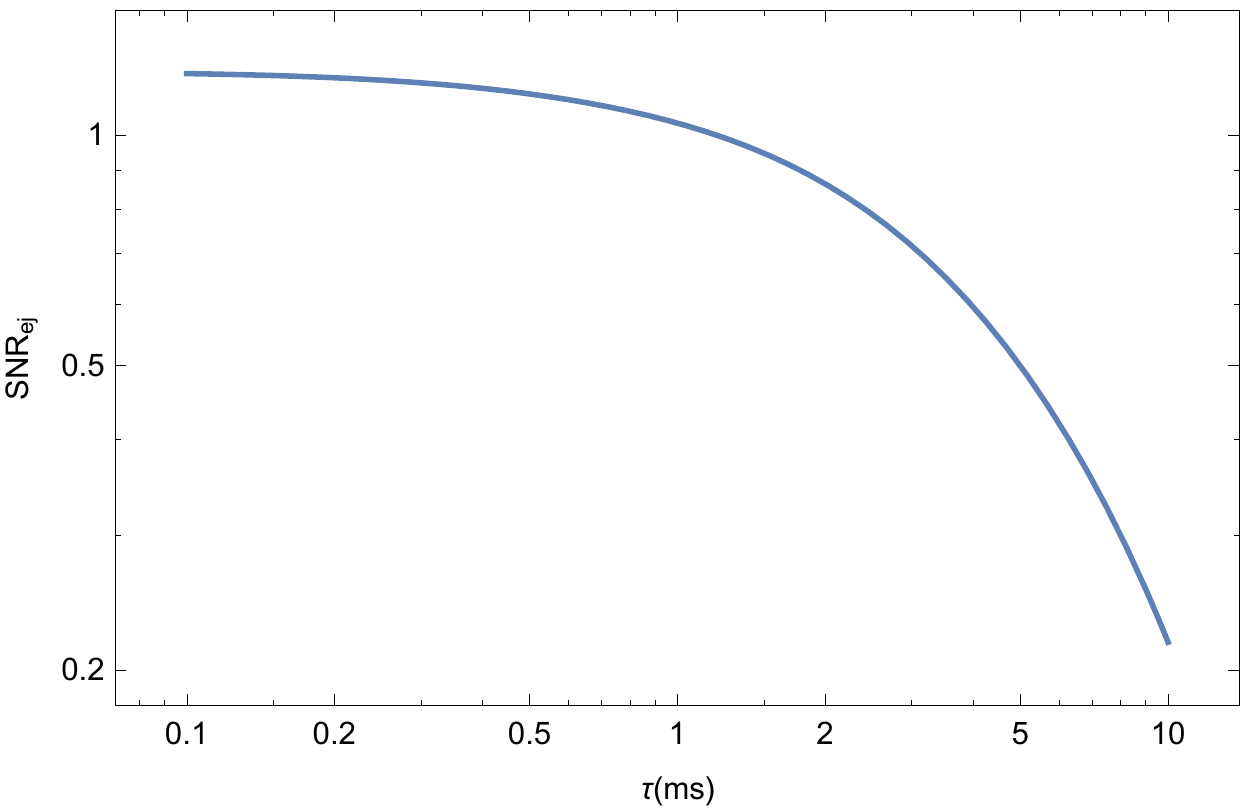}
\caption{SNR  for a binary neutron star system at $50$Mpc, assuming ejecta mass $\Delta M=0.03 M_\odot$ and characteristic ejecta speed to be $0.3$c.}
\label{fig:ej}
\end{figure} 

\bibliography{master}
\end{document}